\documentstyle[11pt,titlepage]{article}  
\titlepage  
 
\begin{document} 
\begin{center} 
{\Large \bf Brownian Motion on a Sphere: Distribution of Solid Angles } 
\vskip 1cm 
{\large M.~M.~G. Krishna${}^{1,2}$, Joseph Samuel${}^{3}$ 
and Supurna Sinha${}^4$} \\ 
${}^1$Department of Chemical Sciences\\
Tata Institute of Fundamental Research, Mumbai 400 005, INDIA.\\
 ${}^2$Department of Biochemistry and Biophysics,\\\ 
University of Pennsylvania School of Medicine,\\ 
Philadelphia, PA 19104, U.S.A.\\ 
${}^3$Raman Research Institute, Bangalore 560 080, INDIA.\\ 
${}^4$307, Sampige Road \#6 Malleswaram, Bangalore 560 003, INDIA. 
\end{center} 
\newline 
\vskip 5 cm 
email:mmg@hxiris.med.upenn.edu, sam@rri.ernet.in

\newpage  
\section*{Abstract} 
We study the diffusion of Brownian particles on the surface 
of a sphere and compute the distribution of solid angles enclosed 
by the diffusing particles. This function 
describes the distribution of  
geometric phases in two state quantum  
systems (or polarised light) undergoing random evolution.
Our results are also relevant to recent  
experiments which observe the Brownian motion of molecules  
on curved surfaces like micelles and biological membranes.
Our theoretical analysis 
agrees well with the results of computer experiments. 
\newpage 
Let a diffusing particle start from the north pole of  
a sphere at time  
$\tau=0$. We join the final position of the particle at  
time $\beta$ to its initial one (the north  
pole) 
by the shorter geodesic. This rule is well defined, unless the  
final position 
is {\it exactly} at the south pole, a zero probability event. 
The path followed 
by the diffusing particle (closed by the geodesic rule) encloses a 
solid angle $\Omega$. The question we address is: At time $\beta$, 
what is the distribution $P^{\beta}(\Omega)$ of solid angles? 
 
 An experimental motivation for this question comes from 
 recent time-resolved fluorescence 
 studies\cite{MMG,MMG1,quitevis95,maiti95}
 on the Brownian motion  
 of rod-like molecules on 
 curved surfaces
 such as micelles and lipid vesicles. 
 The experimentally measured fluorescence anisotropy is affected
 by the curvature of the surface 
 due to geometric phase effects\cite{MMG,Berry,zwanziger90}. 
 The problem 
 of diffusion on curved surfaces reappears in 
 Nuclear Magnetic Resonance (NMR) and  
Electron Spin Resonance (ESR) studies on micelles and lipid
vesicles.
The curvature of the surface affects the spin
relaxation times\cite{walderhaug84,nery86,kuchel96}. 
In biological systems, the translational 
diffusion of solutes bound to various curved surfaces 
influences metabolic rates and transmission rates of chemical
signals\cite{MMG,kuchel96}.
Recently, fluorescence anisotropy decay of molecules 
on curved surfaces has been studied theoretically and 
using Monte Carlo simulations\cite{MMG}.  
 This study shows that the geometric phase 
crucially affects the anisotropy decay if the molecules are
tilted relative to the surface normal. We therefore need
to theoretically evaluate the distribution of geometric phases
or equivalently, the distribution of solid angles in order to
understand diffusion processes on curved surfaces.
 
In this paper we present an answer to the
question posed above and compare our theoretical results 
with Monte Carlo simulations.
The theoretical analysis presented here  
is  based on earlier work \cite{Edwards,C1,C2,C3,BM}. 
In \cite{C1,BM} 
the distribution of solid angles for {\it closed} Brownian 
paths on the sphere is computed. However, closed Brownian paths are a set of measure zero 
among Brownian paths  
and a comparison of the results of \cite{C1,BM} with real or  
computer experiments 
is hampered by poor statistics. We adapt the methods of  
\cite{C1,BM} 
to allow for {\it open} Brownian paths and compute the  
distribution of 
solid angles for such paths, 
closing the path by the geodesic rule 
\cite{Berry} 
in order to make the solid angle a well-defined quantity. 
The qualitative idea behind \cite{Edwards,C1,C2,C3,BM} and the present 
paper is to use a magnetic field as a ``counter",  
to measure the solid angle enclosed 
in a Brownian Motion. 
 
We first illustrate our method by solving a similar problem  
on the plane.  
Let a diffusing particle start 
from the origin $\vec{r}=0$ of the plane at time $\tau=0$. 
Let us suppose that the particle arrives at $\vec{r_f}$  
at time $\beta$. Join $\vec{r_f}$  to the origin by a straight 
line. The open Brownian path ${\vec r}(\tau)$ closed by a straight 
line encloses an area $A$. What is the probability  
distribution $P(A)$ of areas? 
By ``area" we mean the algebraic area, 
including sign. Area enclosed to the left of the  
diffusing particle counts as positive and area to the right 
as negative. This problem has been posed  
and solved \cite{ISIbook} earlier.  
We illustrate our method by  
solving this problem before moving to the problem of  
main interest in this paper.  
 
Let $\{{\vec r}(\tau)=\{x(\tau),y(\tau)\}, 0\leq \tau \leq \beta \,\, , 
{\vec r}(0)=\vec{0}\}$  
be a realization of a Brownian path on the plane which starts at 
the origin. As is well known, Brownian paths are 
distributed according to the Wiener Measure \cite{Wienerref}: 
if ${\cal F}[{\vec r}(\tau)]$ is any functional on paths, the  
expectation value of ${\cal F}$ is given by 
\begin{equation} 
<{\cal F}[{\vec r}(\tau)]>_{\cal W}:={\int {d}{\vec r_{f}} 
\int\ {\cal D} [{\vec r}(\tau)] {\cal F}[{\vec r}(\tau)] 
exp[-\int_0^\beta\{{1\over 2}{d{\vec r}\over{d\tau}}. 
{d{\vec r}\over{d\tau}}{d\tau}\}] 
\over  
\int {d}{\vec r_{f}} 
\int\ {\cal D} [{\vec r}(\tau)] 
exp[-\int_0^\beta\{{1\over 2}{d{\vec r}\over{d\tau}} 
.{d{\vec r}\over{d\tau}}{d\tau}\}] 
} 
\label{Wiener} 
\end{equation} 
 
In Eq. (\ref{Wiener}) the  symbol
${\cal D} [{\vec r}(\tau)] $ denotes a functional  
integral \cite{Feynman} over all  
paths which start at the origin and end at ${\vec r_{f}}$. 
Finally, the endpoint ${\vec r_{f}}$ is also integrated over. 
(We set the diffusion constant equal to 
half.) 
$\beta$ is the time for 
which the diffusion has occured \cite{footnotebeta}. 
Let ${\cal A}[\vec r(\tau)]$ be the  
algebraic area enclosed by the path $\vec r(\tau)$.  
Clearly, the normalized probability distribution of areas 
$P(A)$ is given by  
\begin{equation} 
P(A) := {< \delta({\cal A}[\vec r(\tau)] - A) >}_{\cal W}. 
\label{Defprob} 
\end{equation} 
The expectation value ${\overline{\phi}}$  
of any function $\phi(A)$ of the area  
is given by $\int {P(A)\phi(A)dA}$.  
As is usual in probability theory, we focus on the  
generating function $\tilde{P}(B)$ of  
the distribution $P(A)$ : 
\begin{equation} 
{{\tilde P}(B)}:= {\overline{e^{iBA}}}  
=\int {P(A)e^{iBA}dA}, 
\label{Defpb} 
\end{equation} 
which is simply the Fourier transform of $P(A)$.  
For reasons that will soon be clear, we use the symbol $B$ for the 
Fourier transform  
variable. 
The distribution $P(A)$ can be recovered from its  
generating function by an inverse Fourier transform. 
From Eqs.(\ref{Defprob}) and (\ref{Defpb}) above we find 
\begin{equation} 
{\tilde P(B)} = < e^{iB{\cal A}} >_{\cal W} 
\label{pbw} 
\end{equation} 
 
Let us introduce a fictitious vector potential  
${\vec A}.d{\vec r}=(B/2)(xdy-ydx)$. Clearly $\oint{\vec A}. 
d{\vec r} =B{\cal A}$, where the integral is over the closed 
circuit consisting of the Brownian path closed by the geodesic. 
${\vec A}$  has been 
chosen so that its radial component ${\vec A}. {\vec r}$ vanishes 
and consequently  
the contribution of $\int {\vec A}.d{\vec r}$ along the geodesic 
segment from ${\vec r_{f}}$  to ${\vec 0}$ vanishes. 
It follows that $B{\cal A}$ can be expressed as  
\begin{equation} 
{B{\cal A}}[{\vec r}(\tau)]  
= \int_0^{\beta}{\vec A(\vec r(\tau))}.{d{\vec r(\tau)}\over d\tau} d\tau, 
\label{Flux} 
\end{equation} 
Eqs.(\ref{Wiener}), (\ref{pbw}) 
and (\ref{Flux}) yield  
\begin{equation} 
{\tilde P}(B) ={\int {d}{\vec r_{f}} 
\int\ {\cal D} [{\vec r}(\tau)] 
exp[\int_0^\beta\{-{1\over 2}{d{\vec r}\over{d\tau}}.{d{\vec r} 
\over{d\tau}}{d\tau}\} 
+i\int_0^\beta\{\vec A.{d{\vec r}\over{d\tau}}{d\tau}\}] 
\over{  
\int {d}{\vec r_{f}} 
\int\{\cal D} [{\vec r}(\tau)] 
exp[\int_0^\beta\{-{1\over 2}{d{\vec r}\over{d\tau}}. 
{d{\vec r}\over{d\tau}}{d\tau}\} 
]} 
\label{ratio} 
\end{equation} 
By inspection of Eq.(\ref{ratio}) we arrive at:  
\begin{equation} 
{{\tilde P}(B) = {Y(B) \over Y(0)}}, 
\label{hbt} 
\end{equation} 
where $Y(B)$ is given by  
$Y(B)=\int d{\vec r_{f}} K^B({\vec 0},\vec r_f)$, 
where $K^B({\vec 0},\vec r_f)$ is the quantum amplitude for a 
particle of unit charge and mass to go from the 
origin $\vec 0$ to $\vec r_f$ in imaginary  time $\beta$  
in the presence of a homogenous magnetic field. This 
amplitude can also be expressed as \cite{Feynman} 
\begin{equation} 
K^B({\vec 0},\vec r_f)=\sum_{\{n\}}\exp[-\beta E_n]u^{*}_{n}({\vec 0}) 
u_{n}({\vec r_{f}}) , 
\label{expansion} 
\end{equation} 
where $u_n({\vec r})$  are a complete set of normalised  
eigenstates 
of the Hamiltonian $\hat H=(1/2)(-i{\vec \nabla}-{\vec A})^2$ and $E_n$ are the 
corresponding eigenvalues.  
(Throughout this paper we set $\hbar = c = 1$.)  
We thus arrive at the expression: 
$Y(B) = {\int {d}{\vec r_{f}}\sum_{\{n\}}exp-[{\beta E_{n}}]  
u^{*}_{n}({\vec 0}) 
u_{n}({\vec r_{f}})}$. 
 
Now we demonstrate the utility of Eq.(\ref{hbt})  
by computing the distribution of areas  
for diffusion on a plane. 
The function 
$Y(B)$ for a particle of unit mass and charge  
in a constant magnetic field, 
is easily computed\cite{Landau} from the energies 
$E_{n} = (n+{1 \over 2})B$ and  
the eigenfunctions  
$u_{n}({\vec r_{f}}) = \sqrt{B} \exp{[(-B/4)r_f{}^2]} L_{n}[(B/2)r_f{}^2]$ 
with $L_{n}$, the $n^{th}$ Laguerre polynomial. $r_f$ is the modulus 
of the vector $\vec r_{f}$.  
$$Y(B)  
= {{2 \pi} \over \cosh({{\beta B} \over 2})}.$$ 
From (\ref{hbt}) we find  
${\tilde P}(B) = {  [\cosh({{\beta B}/ 2})]^{-1}}.$ 
Taking the Fourier transform of ${\tilde P}(B)$ by contour  
integration 
we get the result  
${P}(A) = {  [\beta\cosh(\pi A / \beta)]^{-1}}$ as 
derived in \cite{ISIbook}. 
This provides a check on Eq.(\ref{hbt}) and illustrates  
its use.  
 
Let us now address the problem posed at the beginning of this  
paper: 
What is the distribution $P(\Omega)$ of solid angles ($\Omega$)  
enclosed by 
a Brownian particle starting at time $\tau=0$ at the north pole 
${\hat r}_n$ 
of an unit sphere and ending at any other point ${\hat r}_f$  
on the sphere, the 
final point being joined to the initial one by a geodesic. 
 (As stated earlier this rule breaks down only if the final point is  
the south pole, which is a zero probability event). 
Unlike the planar case, 
$P(\Omega)$ is a 
periodic\cite{footnotea} 
function with period $4\pi$. 
The generating function  
${\tilde P}_{g}$ of the distribution of solid angles is given by  
\begin{equation} 
 {\tilde P_{g}}  =  
\int_{0}^{4\pi} {d {\Omega} P(\Omega) 
       e^{i{g\Omega}}  } 
\label{fourierinv} 
\end{equation} 
with $g$ a half integer\cite{footnotec}. 
$P(\Omega)$ is expressed in terms of ${\tilde P} _{g}$ by  
a Fourier series (rather than an integral) with $g$ ranging 
from $-\infty$ to $\infty$ in half integer steps: 
\begin{equation} 
{ P(\Omega) = {1 \over 4\pi}\sum_{g=-\infty}^{\infty} 
       e^{-i{g\Omega}}{{{\tilde P} _{g}}}.} 
\label{fourier}  
\end{equation} 
  
Relation (\ref{hbt}) now takes the form  
\begin{equation} 
{\tilde P}_{g} = {Y_{g} \over Y_{0}}, 
\label{hbt2} 
\end{equation} 
where $Y_g$ is given by  
$Y_{g} = {\int {d}}{\hat r_{f}}\sum_{\{j,m\}} 
exp-[{\beta E^{g}_{j,m}}]  
u^{{*}{g}}_{j,m}({\hat r}_n) 
u^{g}_{j,m}({\hat r_{f}})$ with $u^{g}_{j,m}({\hat r_{f}})$  
the normalised eigenfunctions 
and $E^{g}_{j,m}$ the energy eigenvalues for a quantum particle  
of 
unit charge on a sphere subject to a magnetic field created by  
a monopole of quantized strength $g$ \cite{Dirac} 
at the center of the sphere. The quantum numbers of the 
eigenstates are $j,m$, where 
$j$ is the 
total angular momentum quantum number and $m$ is its 
$z$ component. We choose the vector potential in the form 
$A_\phi=g(1-\cos \theta), A_\theta=0$, which is non singular 
everywhere on the sphere except the south pole.  
Since $A_\theta$ vanishes, (as $A_r$ did in the planar case),  
$\int A$ along the open Brownian path does measure the solid angle  
as defined by the geodesic rule\cite{Berry}.
 
The Hamiltonian for this problem is ${\hat H}^g=-{1\over 2}  
{1\over{\sin\theta}}  
{\partial \over{\partial \theta}} 
 \sin\theta  {\partial \over{\partial \theta}} 
+(-i{\partial \over \partial \phi}-A^g{}_\phi)^2$ 
in standard spherical co-ordinates. 
The wave equation can be separated by writing $\psi(\theta,\phi)=exp(ip_\phi \phi) R(\theta)$. From continuity of $\psi$, it 
follows that $R(0)=0$ if $p_\phi \neq0$. Therefore, 
only those eigenstates for which $p_\phi=0$, 
contribute 
to $Y^g$. These normalised eigenstates are labelled 
by $j$, which ranges from  $|g|$ to $\infty$ in integer steps. 
\begin{equation} 
R^g{}_j(z)=\sqrt{2j+1\over 2} {1\over 2^g} 
(1+z)^g P^{0,2g}{}_{j-g}(z), 
\label{eigenfunctions} 
\end{equation} 
where $z=\cos \theta$ and $P^{a,b}{}_n$ are the Jacobi 
Polynomials \cite{Edmonds}.  
The energy levels of this system are \cite{Coleman,Edmonds}:  
$E^g{}_j={j(j+1)-g^2\over 2}$.Using the integer $n=j-|g|$ to 
label the states (in place of $j$) we find 
\begin{equation} 
{\tilde P}^g=\sum_{n=0}^{\infty} 
 (-1)^n\exp(-[n(n+1)+g(2n+1)]\beta/2) 
{(2n+2g+1)g\over (g+n)(g+n+1)} ,  
\label{Pg} 
\end{equation} 
for $g>0$ and ${\tilde P}^0=1$. 
From (\ref{fourier}) we arrive at \cite{footnoteb} 
\begin{equation} 
 P(\Omega) =(1/4\pi)(1+2\sum_{g=1/2}^{\infty}\cos(\Omega g){\tilde P}_g), 
\label{finalexp} 
\end{equation} 
where $g$ increases in half integer steps. 
The function 
(\ref{finalexp}) is plotted numerically  
for various values of $\beta$ in Fig. $1$.  
The qualitative nature of these plots is easily understood. 
For small values of $\beta$ the particle tends to make small  
excursions and its path encloses solid angles close 
to $0$ or $4\pi$ and consequently 
the plots are peaked around these two values. 
As the available time $\beta$ increases, other values of $\Omega$ 
are also probable and  
the peaks tend to spread and the curves to flatten out.   
Finally in the limit of $\beta \rightarrow \infty $ 
the particle has enough time to enclose all possible solid angles  
with  
equal probability. These plots give the answer to the  
question that was raised in the beginning of the paper.   
  
The analytical solution given above 
was checked
against the results of computer experiments.
The diffusion process was simulated on a  
Silicon Graphics workstation by Monte Carlo methods. 
The simulation methodology was adapted from \cite{MMG} 
(where more details are given) and is
briefly as follows. We start the simulation with  
$10^5$ molecules located at the north pole of a 
sphere with all molecular dipoles aligned parallel 
to the spherical surface 
(perpendicular to the position vectors). 
Let $\hat{V}_{0}$ represent the dipoles at zero 
time, oriented along the laboratory x-axis 
which is perpendicular to the position vector 
$\hat{R}_0$ along the laboratory z-axis.  
These molecules are then subjected to random ``kicks''
 and allowed to diffuse  
over the sphere independent of each other. The diffusion of 
the dipoles on the spherical surface is effected 
by rotation of the position vector $\hat{R}$ 
and the dipole vector $\hat{V}$ by the 
same three dimensional rotation matrix about a 
randomly chosen vector $\hat{n}$ normal to the radial position 
vector $\hat{R}$. In the simulation, the diffusion constant
$D$ is chosen to be $10^{-3}$ per time step and 
the angle of rotation per 
time step (one iteration) is obtained from the corresponding 
probability 
equation for planar diffusion, 
which is valid for very small 
diffusion lengths. The orientation of the 
molecule $\hat{V}$ during diffusion is used as an 
index of the solid angle swept out by the Brownian 
path. We numerically evaluate this for each molecule 
applying the geodesic rule.
Let $\hat{V}_\beta$ and $\hat{R}_\beta$ 
represent the dipole and the radial vectors 
at time $\beta$. To determine the orientation 
of $\hat{V}_\beta $, $\hat{V}_\beta$ is 
transformed to 
$\hat{V'}_\beta$ using a three dimensional rotation 
matrix with an angle of rotation (less than $\pi$) 
determined by  
$cos^{-1}(\hat{R}_\beta.\hat{R}_0)$. 
The angle between $\hat{V'}_\beta$ and $\hat{V}_0$
is the solid angle enclosed by the Brownian path of the 
diffusing molecule at time $\beta$. 
Similarly, the solid angles are 
calculated for all the $10^5$ dipoles at different times. 
These are sorted into 360 
different bins, each corresponding to one degree 
increment between $0$ and $2\pi$ and are plotted in Figure 2. 
 
 Since the  orientation of the molecule 
is only determined modulo $2\pi$ 
in the simulation, $\Omega$ is only measured modulo $2\pi$ 
and not modulo $4\pi$. So in effect the function computed 
by the simulation is $Q(\Omega)=P(\Omega)+P(\Omega+2\pi)$. 
From (\ref{finalexp}) it is clear that the half integer 
values of $g$ cancel out and the integer values are doubled. 
Our theoretical expression for $Q(\Omega)$ is therefore  
\begin{equation} 
 Q(\Omega) =(1/2\pi)(1+2\sum_{g=1}^{\infty}\cos(\Omega g){\tilde P}_g), 
\label{finalexp1} 
\end{equation} 
where $g$ now increases in {\it integer} steps. 
The results of the numerical simulation and the function 
$Q(\Omega)$ (with suitable cutoffs on $g$ and $n$) 
given above are 
plotted in Figure 2 for different values of $\beta$. 
The agreement between the simulations and the theory 
is excellent and the  fractional
deviation is of order  
$1/\sqrt{N}$ where $N$ is the number of 
molecules in a bin.

In computing the distribution of solid angles, 
we have chosen all the molecules to be initially   
at the north pole (${\hat r}_i={\hat r}_n$) 
and integrated over the final  
point ${\hat r}_f$ with a uniform 
weight, to match with the procedure followed in our simulation. 
In an experimental situation, where the diffusing 
particles are excited and observed with external probes, it may 
be necessary to consider a weighted average over the initial 
and final positions of the diffusing particles \cite{MMG}.  
Our analysis is 
easily adapted to take this into account. One just writes 
\begin{equation} 
Y^g=\int d{\hat r}_f 
 w_f({\hat r}_f) d{\hat r}_i w_i({\hat r}_i ) 
 K^g({\hat r}_i, {\hat r}_f). 
\label{weight} 
\end{equation} 
This leads to $Y^g=\sum_n w^g{}^*{}_i w^g{}_f 
 \exp[-\beta E^g_n/2]$, where  
 $w^g{}_n=\int d{\hat r} w({\hat r}) u^g_n({\hat r})$ and the sum is  
 over all eigenstates. In this paper we have set $w_i({\hat r}_i)= 
 \delta^2({\hat r}_i-{\hat r}_n)$ and $w_f({\hat r}_f)=1$. 
  
In the problem of diffusion of fluorescent  
molecules, the tangential  
component of the orientation of the molecule determines the 
solid angle enclosed by the diffusing particle only modulo $2\pi$. 
This is because, in  
parallel transport of a vector, the vector rotates by an angle 
{\it equal} to the solid angle enclosed by the curve. As  
Pancharatnam showed \cite{Berry},  
a quantum two state system transported on the Poincar{\'e} 
sphere picks up a geometric phase equal to {\it half} the  
enclosed solid angle. This measures the solid angle modulo  
$4\pi$. The distribution $P(\Omega)$ computed in the body of the
text describes the distribution of geometric phases. 
However, since our simulations 
were adapted from \cite{MMG}, the distribution measured in the 
simulation is $Q(\Omega)$ 
and not $P(\Omega)$. 

In \cite{C1,BM}, the generating function 
for the distribution of solid
angles for {\it closed} Brownian paths involved
{\it only} the energy eigenvalues of the Hamiltonian.
In contrast, in the present case of {\it open} Brownian paths
(closed by the geodesic rule), 
one needs to use {\it both} the
eigenvalues as well as the {\it eigenfunctions} 
of the Hamiltonian.
This is the main difference between \cite{C1,BM} and the present
theoretical analysis.

We have computed the distribution of solid angles for Brownian 
motion on the sphere. This calculation has also been checked  
against computer experiments. In simple geometries like the  
surface of a sphere, an analytical treatment is possible.  
In more complicated situations, one is forced to rely on 
computer simulations. Monte Carlo simulations can be used
in two distinct roles. In the present problem, where
an analytical treatment is possible, simulations can be viewed
as a computer experiment against which the theory can be
tested. Such experiments are much easier 
to control than real experiments.
In cases where an analytical treatment
is not possible, the simulations serve as a substitute for the
theory, to be checked against real experiments.
The agreement between theory and  
computer experiments for Brownian motion on the sphere 
also serves as a check on the computer simulations so that 
these simulations can be confidently used in  
geometries which cannot be treated analytically.

{\it Acknowledgements}: It is a pleasure to thank R. Nityananda 
for discussions on this subject and S. Ramasubramanian for drawing 
our attention to Ref. \cite{ISIbook} and Y. Hatwalne for a critical
reading of the manuscript. JS and SS thank N. Kumar for
raising the question of the distribution of solid angles
on a sphere. MMGK thanks N. Periasamy and S.W. 
Englander for their encouragement. We thank Alain Comtet for
drawing our attention to some relevant references.

\newpage 
\section*{Figure Captions} 
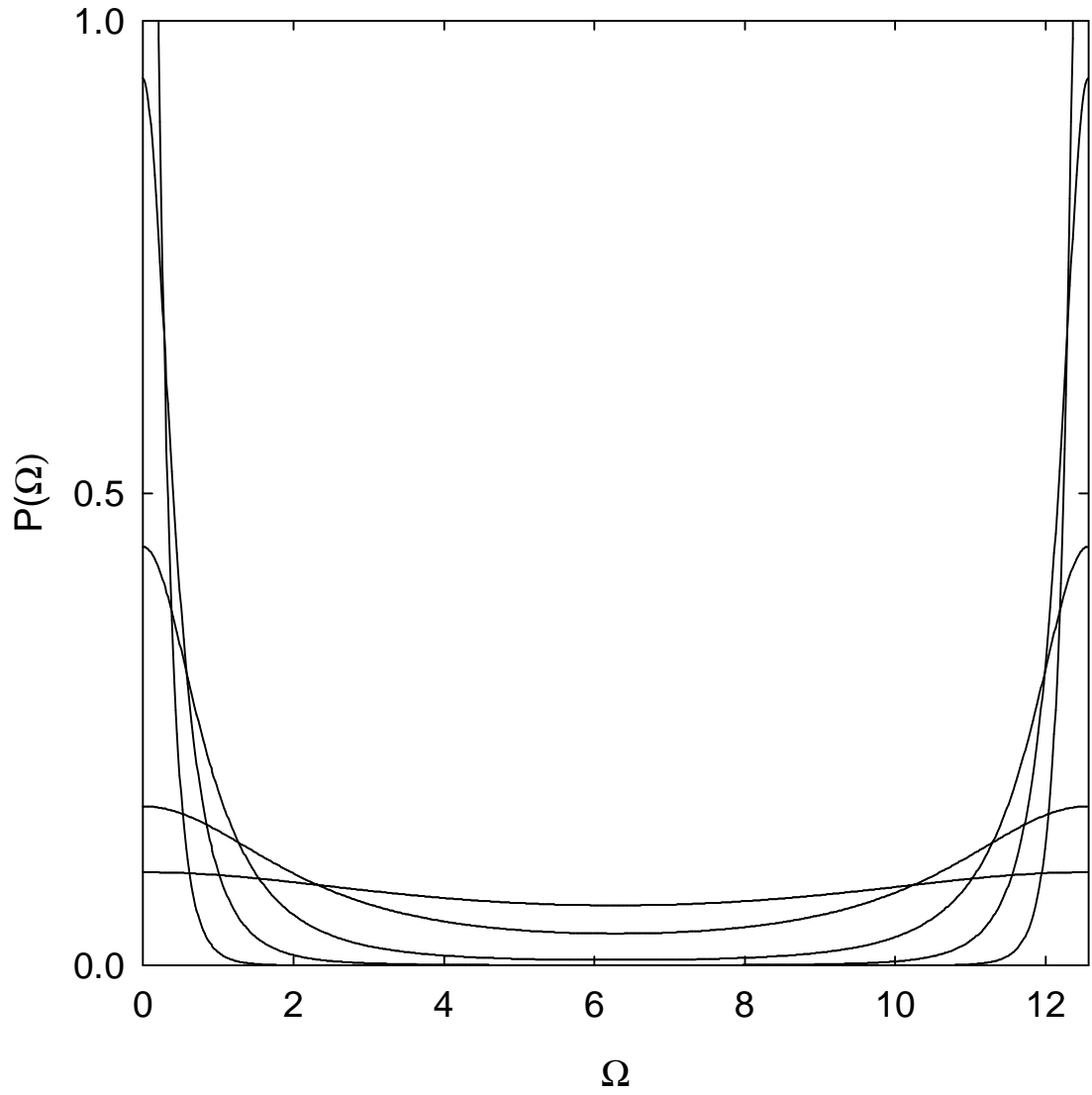
\noindent Fig. 1: The solid angle distributions $P(\Omega)$ 
computed from Eq. \ref{finalexp} at five different values 
of $\beta$: 0.5, 1, 2, 5 and 10. The flatter curves correspond
to higher $\beta$.
\newline 
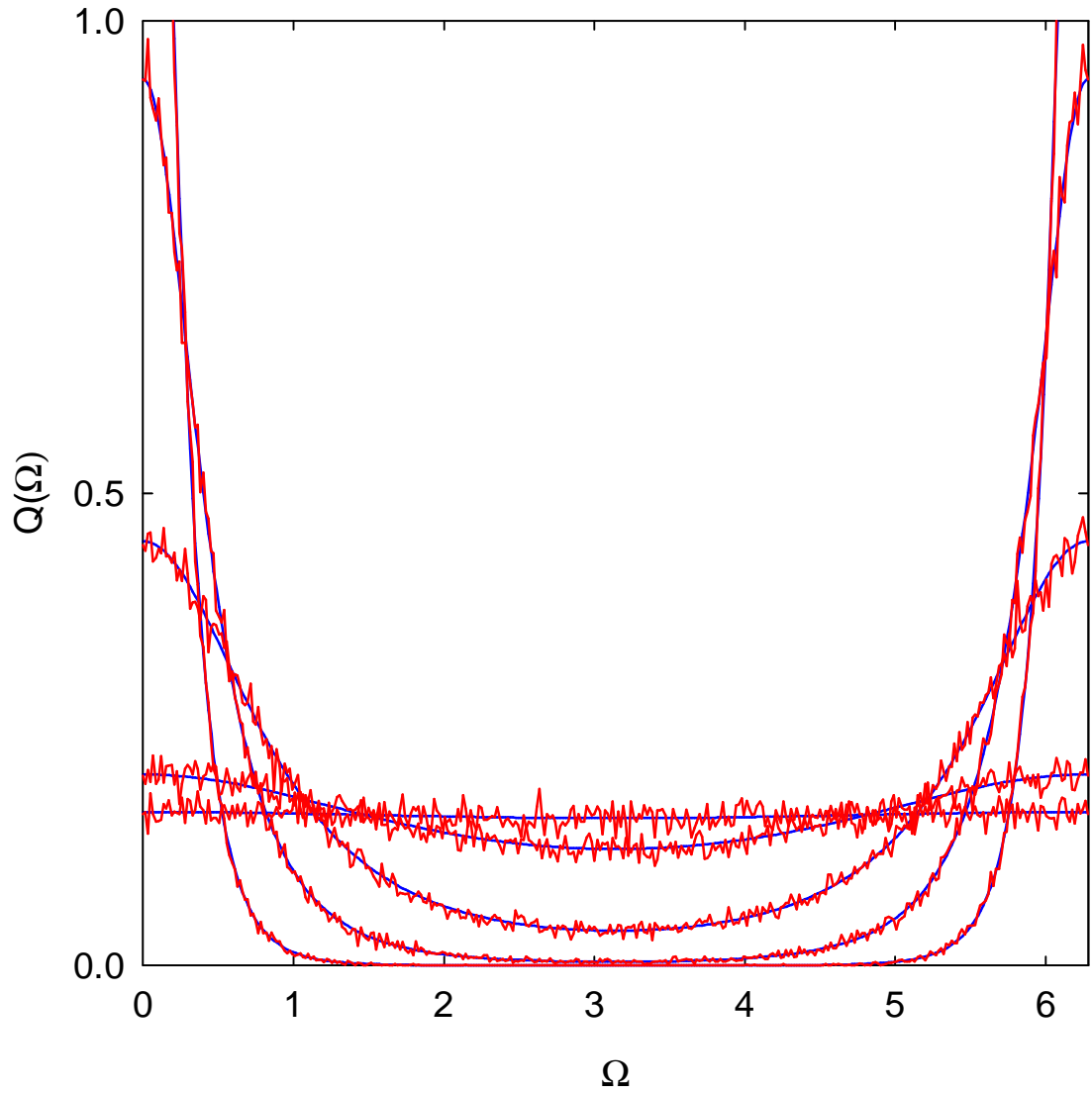
\noindent Fig. 2: Comparision between solid angle 
distributions $Q(\Omega)$ from Monte Carlo 
simulations (red line) and the theory (Eq. \ref{finalexp1})
(blue line). The plot 
shows the distributions at five different values of 
$\beta$: 0.5, 1, 2, 5 and 10.

\begin{thebibliography}{99}
\bibitem{footnotebeta} 
{In real applications, 
$\beta$ is equal to $2Dt$ in the case
of a plane and is equal to $2Dt/R^2$ in the case of a sphere,
where $D$ is the diffusion 
coefficient on the surface, $R$ is the radius  
and $t$ is the time for which the particle has diffused.
In the text we have chosen units so that $R=1,
D=1/2$ then $\beta$ is equal to the time.}
\bibitem{footnotea} 
{A path that encloses a solid angle $\Omega$  
to its left also  
encloses a solid angle of $4\pi - \Omega$ to its right. Using  
the symmetry $P(\Omega) = P(-\Omega)$ we conclude that  
$P(\Omega)$ is periodic with period $4\pi$.}
\bibitem{footnotec} 
Note that the definition of $g$ here is {\it not} the same 
as in ref. \cite{BM}, which contains an error: 
in the equation 
before (11) $g$ should be replaced by $g/2$.  
This error propagates to equations (11,12) and Figure 1 but not 
to the subsequent discussion.
\end{thebibliography}
\end{document}